\documentclass[11pt,a4paper]{article}
\pdfoutput=1
\usepackage{jinstpub}
\usepackage{graphicx}
\usepackage{epstopdf}
\usepackage{dcolumn}
\usepackage{bm}
\usepackage{siunitx}
\usepackage{booktabs}
\usepackage[normalem]{ulem} 
\usepackage{threeparttable} 
\usepackage{tabularx}
\usepackage{siunitx}
\usepackage{adjustbox}
\usepackage{verbatim}

\usepackage[colorinlistoftodos]{todonotes}

\begin{document}

\title{\boldmath Direct comparison of high voltage breakdown measurements in liquid argon and liquid xenon}

\author[a,b,c,d,1,2]{L.~Tvrznikova, \note{Corresponding author}\note{Now at: Waymo LLC, Mountain View, CA 94043, U.S.A.}}
\author[b,c]{E.P.~Bernard,}
\author[c]{S.~Kravitz,}
\author[b,c,3]{K.~O'Sullivan, \note{Now at: Grammarly Inc., San Francisco, CA 94104, U.S.A.}}
\author[b]{G.~Richardson,}
\author[b,c]{Q.~Riffard,}
\author[c]{W.L.~Waldron,}
\author[b]{J.~Watson,}
\author[b,c]{D.N.~McKinsey}

\affiliation[a]{Yale University, Department of Physics, 217 Prospect St., New Haven, CT 06511, U.S.A.} 
\affiliation[b]{University of California Berkeley, Department of Physics, 366 LeConte Hall, Berkeley, CA 94720, U.S.A.} 
\affiliation[c]{Lawrence Berkeley National Laboratory, 1 Cyclotron Rd., Berkeley, CA 94720, U.S.A.}
\affiliation[d]{Lawrence Livermore National Laboratory, 7000 East Ave, Livermore, CA 94550, U.S.A.}

\emailAdd{lucie.tvrznikova.phd@gmail.com}

\abstract{
As noble liquid time projection chambers grow in size their high voltage requirements increase, and detailed, reproducible studies of dielectric breakdown and the onset of electroluminescence are needed to inform their design. The Xenon Breakdown Apparatus (XeBrA) is a 5-liter cryogenic chamber built to characterize the DC high voltage breakdown behavior of liquid xenon and liquid argon. Electrodes with areas up to 33~cm$^2$ were tested while varying the cathode-anode separation from 1 to 6~mm with a voltage difference up to 75~kV. A power-law relationship between breakdown field and electrode area was observed. The breakdown behavior of liquid argon and liquid xenon within the same experimental apparatus was comparable.}

\arxivnumber{1908.06888}
\maketitle
\flushbottom

\section{Introduction}

The goal of the Xenon Breakdown Apparatus (XeBrA) at the Lawrence Berkeley National Laboratory (LBNL) is to characterize the high voltage (HV) breakdown behavior in liquid argon (LAr) and liquid xenon (LXe). Its characterization improves our understanding of DC dielectric breakdown and also informs the design of the next-generation of noble liquid time projection chambers (TPCs), such as DarkSide20k~\cite{Aalseth:2017fik}, DUNE~\cite{Abi:2018dnh}, DARWIN~\cite{Aalbers:2016jon}, and nEXO~\cite{Albert:2017hjq}. Dielectric breakdown in a two-phase TPC is a concern because it can damage detector hardware, and detector performance can be compromised by spurious light and charge emission at lower voltages before complete breakdown occurs.

The original measurements of LAr breakdown were made over micrometer gap distances~\cite{swan1960influence,swan_LAr}, but experiments at larger scales were not able to reach the fields of these first studies. Later measurements in LAr at millimeter scales revealed that the threshold for dielectric breakdown decreases as the surface area of the electrodes is increased ~\cite{Acciarri:2014ica,Auger:2015xlo}. This behavior is likely also present in LXe, but comprehensive studies of dielectric breakdown behavior for large area electrodes in LXe have not been published. 

Dielectric breakdown has been of interest for many decades in various other materials due to their application as insulators in electric power generation and transmission. Here, the focus is on breakdowns under DC fields in noble liquid dielectrics. A universally agreed-upon factor affecting DC breakdown behavior is the stressed area effect~\cite{weber1956nitrogen,Auger:2015xlo,weber1955areaEffect,KAWASHIMA1974217,goshima1995nitrogen,GERHOLD1994579,Gerhold1989,Acciarri:2014ica,goshime1995weibull,Gerhold_LiquidHelium,toya1981}. Some studies also report a decreasing dielectric strength with an increase of the stressed liquid volume~\cite{GERHOLD1994579,goshime1995weibull,Gerhold_LiquidHelium,goshima1995nitrogen}. However, unless the tested geometries are very different, the area and volume effects are hard to distinguish. 

Even with this common understanding, a first-principles theoretical description of the dependence of breakdown on the electrode has still not been developed. The weakest-link Weibull model provides an empirical explanation of the statistical origins of the area effect and is discussed in section~\ref{sec:weak_link}. The physics of the initiation of dielectric breakdown and the succeeding discharge evolution within liquids remains an active subject of research~\cite{plasma_physics}. Other than the electrode area, many other physical parameters have been observed to affect dielectric breakdown:
\begin{itemize}
    \item \textbf{Surface finish:} Electrodes with a mirror finish have been shown to sustain higher electric fields in liquid helium (LHe)~\cite{GERHOLD1994579,Gerhold1989} and in liquid nitrogen (LN)~\cite{goshima1995nitrogen} than less polished electrodes. Electron emission from electrode surfaces is likely an initiator of dielectric breakdown; it has been shown that emission rates from cathodic stainless steel wires into LXe are greatly reduced by acid passivation and electropolishing processes \cite{TOMAS201849}. 
    \item \textbf{Material:} Electrode materials, as well as surface oxidation, have been shown to affect dielectric breakdown in LAr~\cite{swan_LAr} and LHe~\cite{GERHOLD1994579,Gerhold1989, Yoshino1982}. 
    \item \textbf{Pressure and temperature:} Breakdown fields increased with increasing pressure in LN~\cite{KAWASHIMA1974217, Yoshino1982} and LHe~\cite{Gerhold1989,long2006high}. However, some earlier studies in LN saw a decrease in breakdown fields with increasing temperature, ~\cite{KAWASHIMA1974217} which correlates directly with pressure.
    \item \textbf{Purity:} Impurities lowered breakdown fields in transformer oil~\cite{ikeda_movingOil} and LHe~\cite{Gerhold1989, Yoshino1982}. Conversely, in LAr, the presence of impurities increased breakdown fields~\cite{swan_LAr,Acciarri:2014ica,blatter2014experimental}.
    \item \textbf{Polarity:} In LAr~\cite{Acciarri:2014ica,blatter2014experimental}, LN~\cite{goshima1995nitrogen}, and LHe~\cite{Gerhold1989,schwenterly1974}, research has shown that sparks during dielectric breakdown are initialized on the cathode surface and are affected by space-charge effects on the cathode surface and in the stressed volume. However, the formation of filamentary discharges, known as streamers, has been shown to be preferentially initiated at the anode in a wide range of other liquids~\cite{streamer}.
    \item \textbf{Capacitance:} References~\cite{cross1982,mazurek1987} have observed that adding parallel capacitance to a pair of electrodes separated by vacuum or transformer oil lowered the threshold of dielectric breakdown and increased damage to the electrodes. This is attributed to the increased stored energy accessible to a growing discharge.
    \item \textbf{Circulation speed:} The breakdown field in transformer oil has been shown to depend on fluid circulation velocity~\cite{ikeda_movingOil}.
    \item \textbf{Conditioning:} Conditioning is frequently used in vacuum to increase the breakdown fields of surfaces, particularly with impulse voltage~\cite{toya1981,ballat,yasuokaelectrode}. It is similarly used in wire chambers and transformer oil~\cite{ikeda_movingOil}. Applying high electric fields to critical regions serves to ablate impurities and surface irregularities that might initiate breakdown. This is only productive if the conditioning discharges do not damage the electrode surfaces. The utility of conditioning electrode surfaces is unclear in LHe~\cite{Gerhold1989,Olivier_Helium,Coletti1982} and unreported for other noble liquids.
\end{itemize}

\subsection{The Weibull weakest-link model as applied to surface-initiated dielectric breakdown} \label{sec:weak_link}

Consider an electrode surface that is subdivided into many independent elements, each of which must withstand an applied electric field $E$ for the entire electrode to avoid breakdown. Weibull recognized that dielectric breakdown could be described with a weakest-link model, in which the failure of an object composed of many independent elements is determined by the failure of its weakest component~\cite{Weibull1939}. In such objects, if the survival probabilities of the individual elements follow power-law distributions, the scaling behavior of the survival probability of the composed object can be predicted~\cite{Weibull1939, Weibull1951}. The statistical basis for weakest-link models is called extreme-value theory.

Suppose that the probability $S_0$ that a surface element of area $\alpha$ can withstand an electric field $E$ follows the power law 
\begin{equation}
S_0(E) = \mathrm{exp}\left[-\left(\frac{E}{\epsilon}\right)^{k}\right]\label{eq:S_element}
\end{equation}
where $\epsilon$ is a reference field that sets the overall scale of the distribution. 
The probability that an electrode composed of $A/\alpha$ area elements, each independent and characterized by $S_0$, can withstand an electric field $E$ is then
\begin{equation}
S(E, A) = 1 - F(E,A) = \left\{ \mathrm{exp}\left[-\left(\frac{E}{\epsilon}\right)^{k}\right] \right\}^{A/\alpha} = \mathrm{exp}\left[-\frac{A}{\alpha}\left(\frac{E}{\epsilon}\right)^{k}\right]\label{eq:S_notime}
\end{equation}
where $F$ is the probability of failure of the electrode. 
Differentiating $F$ with respect to $E$ gives the Weibull distribution of the failure probability as a function of the field:
\begin{equation}
g(E)\Big\rvert_{A} = \frac{k}{\epsilon}\frac{A}{\alpha}\left(\frac{E}{\epsilon}\right)^{k-1}\mathrm{exp}\left[-\frac{A}{\alpha}\left(\frac{E}{\epsilon}\right)^{k}\right] = \frac{k}{\lambda}\left(\frac{E}{\lambda}\right)^{k-1}\mathrm{exp}\left[-\left(\frac{E}{\lambda}\right)^{k}\right]\label{eq:weibull_notime} 
\end{equation}
where $k$ and $\lambda = \epsilon \left(\alpha/A\right)^{1/k}$ are the shape and scale parameters of the distribution. 

The scaling of breakdown distributions with area can be understood through equations (\ref{eq:S_notime}) and (\ref{eq:weibull_notime}). If electrodes of varying area are tested, while the remaining experimental conditions (encoded in $\epsilon$, $\alpha$, and $k$) are fixed, then $S(E,A)$ and $g(E)\big\rvert_{A}$ are functions of only the electric field $E$ and electrode area $A$. Each tested area will result in a distribution of breakdown fields with a median field value $E_{m}(A)$. The median values lie on a line specified by 
\begin{equation}
S(E_m(A), A) = \frac{1}{2}, \label{eq:S_median}
\end{equation}
which leads to 
\begin{equation}
E_m(A) = \epsilon\left(\alpha \ln{2}\right)^{1/k}A^{-1/k} = C \cdot A^{-1/k} = C \cdot A^{-b} \label{eq:AreaScalingGeneric}
\end{equation}
where $b=1/k$ is the area scaling exponent. Other features such as the mean and mode of $g(E)\big\rvert_{A}$ result from setting $S(E,A)$ to other values in equation (\ref{eq:S_median}). These also follow the scaling of equation (\ref{eq:AreaScalingGeneric}), which is a consequence of holding $S(E,A)$ constant.

Equation~(\ref{eq:AreaScalingGeneric}) enables prediction of the breakdown behavior of electrodes with different areas from the Weibull distribution obtained from the breakdown distribution of a single electrode. This approach has been taken in many publications~\cite{weber1956nitrogen,goshime1995weibull,goshima_LN,haywaka_LN,gerhold_LN,8124618,Gerhold_LiquidHelium,1996812,suehiro1996SuperfluidHe,alumina_breakdown,Ulhaq_polymer,Hill_breakdown,choulkov} and most recently confirmed in~\cite{Acciarri:2014ica} for LAr.

The Weibull distribution of equation~(\ref{eq:weibull_notime}) results from the power-law scaling form chosen in equation~(\ref{eq:S_element}). This distribution is one of three that satisfy the requirement that the distribution remains unchanged (except for scale) when the number of elements that may fail is varied \cite{FisherTippett1928, Gyorgyi2010,Coles2001,Kotz2000}. Two other distributions (Gumbel and Fr\'echet) are also applicable to weakest-link theory when modeling distributions in which the tail of events surviving to high stress decays relatively slowly~\cite{NevesAlves2008}.

The experiments described here consist of exposing electrode surfaces to electric fields that increase linearly in time. The effect of ramp rate on the DC breakdown field distribution in noble liquids is uncertain and mostly unexplored. In LAr, the MicroBooNE collaboration found no difference in the breakdown distributions obtained from testing a 1~mm gap under DC fields at 50~V/s and 250~V/s ramp rates~\cite{Acciarri:2014ica}. In normal LHe, increasing the period between ramped AC breakdown tests from 1 to 8 minutes resulted in a 30\% increase in average breakdown field across a 0.4~mm gap between 25~mm radius electrodes~\cite{Coletti1982}. The effects of varying the ramping rate are unreported in LHe except for a measurement of ramped AC field across a 4.5~mm gap, which finds a slight decrease in breakdown field at longer timescales~\cite{Gerhold1998Properties}. Reference~\cite{Hill_breakdown} provides a general framework for incorporating timescale effects into the Weibull analysis of dielectric breakdown. 

\section{Experimental setup}

\subsection{XeBrA design}

\begin{figure}
\begin{center}
\includegraphics[width=115mm]{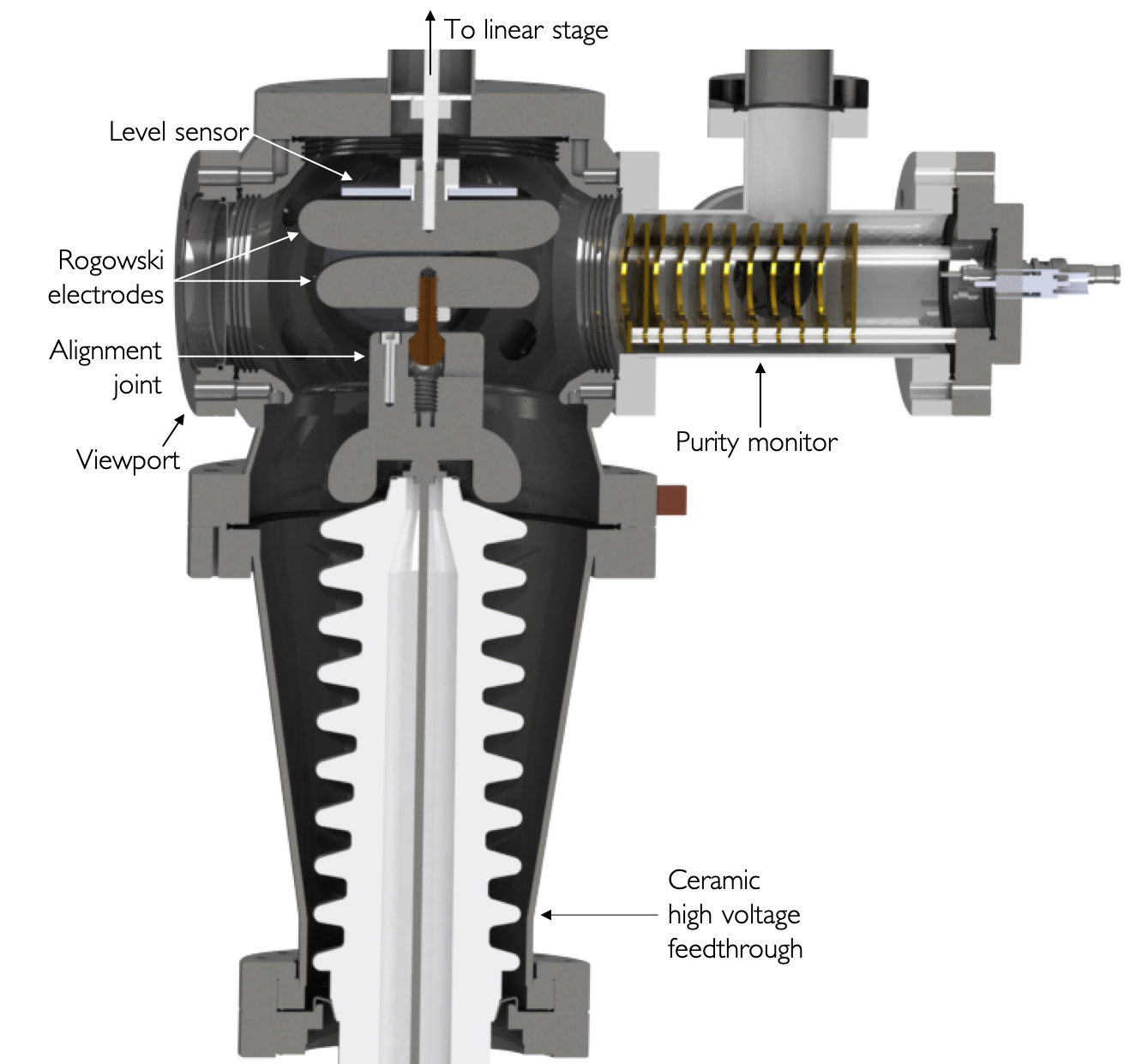}
\caption{A CAD rendering of the active volume of XeBrA. A liquid level sensor is fastened to the top electrode. The cathode (the bottom electrode) is connected to the alumina ceramic HV feedthrough, and the grounded anode is on top, attached to the linear translation stage (not shown). The PMT is obscured by the electrodes in this image.}
\label{fig:CAD_model}
\end{center} 
\end{figure}

\begin{figure}
\begin{center}
\includegraphics[width=150mm]{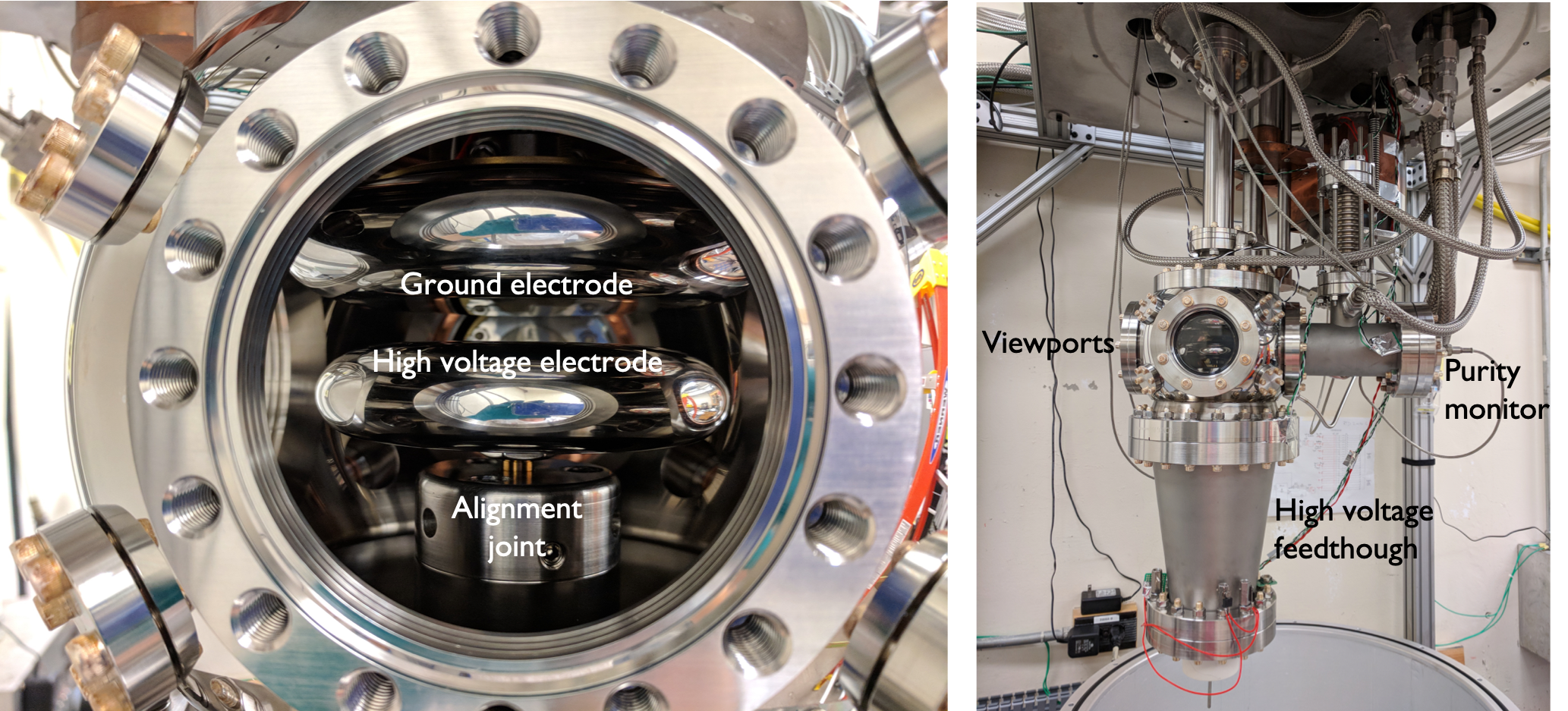}
\caption{Left: image of the assembly inside the central chamber seen through a viewport. Right: view of the xenon volume before the application of superinsulation with the purity monitor attached to the right of the central chamber. The PMT is attached at the back.}
\label{fig:hardware}
\end{center} 
\end{figure}

XeBrA can be filled with either LAr or LXe, making it the first apparatus to enable a direct comparison of HV behavior between these noble liquids. The initial studies in XeBrA were focused on exploring the area effect. The largest reported stressed area in the literature in LAr is only 3~cm$^2$, limited by the geometry of the electrodes. To overcome this limitation, XeBrA utilizes axially symmetric Rogowski electrodes~\cite{rogowski,Rogowski1923} optimized for field uniformity over large areas, providing more relevant insight for large experiments. The CAD rendering of the apparatus is shown in figure~\ref{fig:CAD_model}, and photographs of the apparatus are shown in figure~\ref{fig:hardware}. The top electrode was grounded and attached to a linear translation stage that changed the vertical separation of the electrodes from 1~to 6~mm in this work, corresponding to stressed areas of 11.2 to 32.6~cm$^2$. As the vertical separation between the electrodes increases, so does the stressed electrode area, rising to 58~cm$^2$ at a 10~mm separation. This is because the boundary defining the stressed electrode area grows with the increasing separation of electrodes, which is a feature of their design. Refer to section~\ref{sec:exp-procedure} for details. 

\begin{figure}
\begin{center}
\includegraphics[width=150mm]{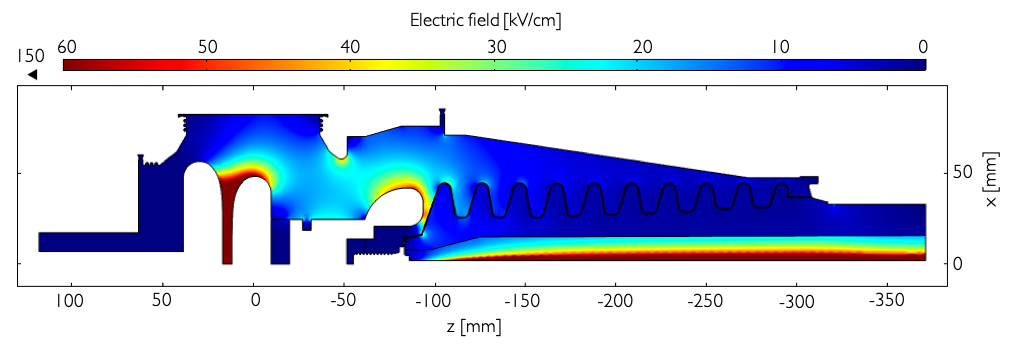}
\caption{Results of an electric field simulation of LXe with -75~kV on the cathode (bottom electrode, toward the right of the figure). Maximum field occurs between the two electrode faces. At 5~mm electrode separation, the maximum field between the electrode faces is 150~kV/cm, while the second-highest field in the apparatus (54~kV/cm) is at the rounded part of the alignment joint.}
\label{fig:field_model}
\end{center} 
\end{figure}

The HV electrode was attached to an alumina ceramic feedthrough\footnote{CeramTec, model 6722-01-CF} rated to 100~kV and 6.5~A DC in vacuum. An alignment joint served as an HV connection between the electrode and the feedthrough, with a ball spindle located near the top of the joint allowing for angular adjustment of the lower electrode. Pressure-sensitive recording film was used to align the lower electrode parallel to the upper electrode within the apparatus after assembly. This ensured uniformity of the electric field between the electrodes. The simulation of the electric field inside the active volume for a 5~mm electrode separation is shown in figure~\ref{fig:field_model}. The DC dielectric constant of LXe was modeled as 1.85, which is typical of measured values in the scientific literature~\cite{doi:10.1063/1.1724850,doi:10.1139/p70-033,doi:10.1142/5113,SAWADA2003449}. The dielectric constant of alumina ceramic was modeled as 9.4. 

The HV power supply\footnote{Spellman, model SL100N10/CMS/LL20} was limited to -75~kV during operations with a maximum current of 10~$\mu$A DC. The power supply could be controlled either manually or remotely through a \textsc{LabView} interface. A 956~k$\Omega$ resistor installed between the power supply and the apparatus limited the peak current and minimized the stored energy available to breakdowns. Polyethylene cables\footnote{Dielectric Sciences, model 2121} connected the power supply, the limiting resistor, and the ceramic feedthrough. The total length of the cable between the limiting resistor and the feedthrough was 2.1~m. The anode was soft-grounded through a 1 k$\Omega$ resistor in series with anti-parallel diodes. Following the resistor was instrumentation to detect current flow that may precede complete breakdown; transient currents were sensed by capacitively coupling the point following the resistor to a charge-sensitive amplifier and to a picoammeter to sense DC leakage currents.

Additionally, a photomultiplier tube (PMT) and a purity monitor were installed directly adjacent to the central chamber, while two of the 4.5~in ConFlat ports were used as viewports. The purity monitor measured the lifetime of electrons as they drift through the noble liquid, which directly relates to the liquid purity level. A purity measurement was desired since prior work in LAr suggests that breakdown strength depends on the presence of impurities in the liquid, with HV breakdown occurring at higher fields with higher levels of electronegative contaminants~\cite{Acciarri:2014ica,blatter2014experimental}. Very low impurity levels are desired in XeBrA because the results are intended to inform direct dark matter and neutrino experiments, which typically operate with very long electron lifetimes of $\mathcal{O}(1\mathrm{~ ms})$. The purity monitor design was similar to the instruments developed for the ICARUS experiment~\cite{bettini1991pm_initial,carugno1990pm_formula}. Details about the design, development, and operating principles of the XeBrA purity monitor can be found in~\cite{Tvrznikova:2018pcz}.

The experimental chamber was wrapped in multi-layer insulation and contained in a vacuum vessel. A cast-epoxy HV feedthrough\footnote{Isolation Products, model D-102-10} rated to 100~kV DC terminated the HV cable from the limiting resistor and provided an HV terminal inside the vacuum vessel. The apparatus was cooled through a heat exchanger attached to the cold head of a pulse tube refrigerator, and the temperature was controlled by a heater on the heat exchanger governed by a feedback loop. Two copper thermal links each conducted $\sim$1.5~W of heat to the cold head from flanges below the central chamber. This sub-cooled the liquid between the electrodes and around the ceramic feedthrough, limiting bubble formation. 

A dedicated gas recirculation and purification system was designed and built for the experiment. A custom oil-free scroll pump\footnote{Air Squared, model P15H22N4.25} continuously circulated gas through a heated zirconium getter\footnote{SAES MonoTorr, model PS4-MT15-R1}, which cleaned the gas of non-noble impurities. The xenon gas was also cleaned of water and oxygen by in-line chemical purifiers\footnote{Matheson Tri-Gas, model SEQPURILOMT1} before the initial gas condensation. The system was instrumented with pressure and temperature transducers and operated through a programmable logic controller (PLC).

\subsection{Experimental procedure and data collection}\label{sec:exp-procedure}

Data presented are from one LAr data acquisition (Run 1) and two LXe data acquisitions (Runs 2 and 3). Data collection started in May 2018 and continued intermittently until September 2018. During acquisition, the HV was ramped automatically by the PLC in 100 V increments. Three ramp rates were used, as shown in table~\ref{tab:Summary-of-detector} along with other parameters relevant to the experiment. The voltage ramp increased until a breakdown, after which the cathode was held at ground for 10-20~seconds prior to starting another ramp. Tens to hundreds of breakdown events were measured at each electrode separation. The electrode separation was calibrated by zeroing the caliper on the linear translation stage as the electrodes touched faces.

\begin{table}
\begin{centering}
\begin{tabular}{lrl}
\hline 
Parameter & Value & Unit\tabularnewline
\hline 
\hline 
Electrode material & 303 SS & \tabularnewline
Cathode surface finish (Ra) & $0.05$ & $\mu$m\tabularnewline
Anode surface finish (Ra) & $0.07$ & $\mu$m\tabularnewline
Power supply limit & -75 & kV\tabularnewline
Trip current & 10 & $\mu$A\tabularnewline
Cable capacitance & 203 & pF\tabularnewline
Ramp speed 1 & 30-60 & kV/min\tabularnewline
Ramp speed 2 & 12 & kV/min\tabularnewline
Ramp speed 3 & 6 & kV/min\tabularnewline
Wait period & 10-20 & s\tabularnewline
\hline 
\end{tabular}
\par
\end{centering}
\caption[Summary of apparatus parameters]{Summary of apparatus parameters pertaining to data collection. Cable capacitance incorporates the length of the cable between the series resistor and the cathode.
\label{tab:Summary-of-detector}}
\end{table}

\begin{figure}
\begin{center}
\includegraphics[width=150mm]{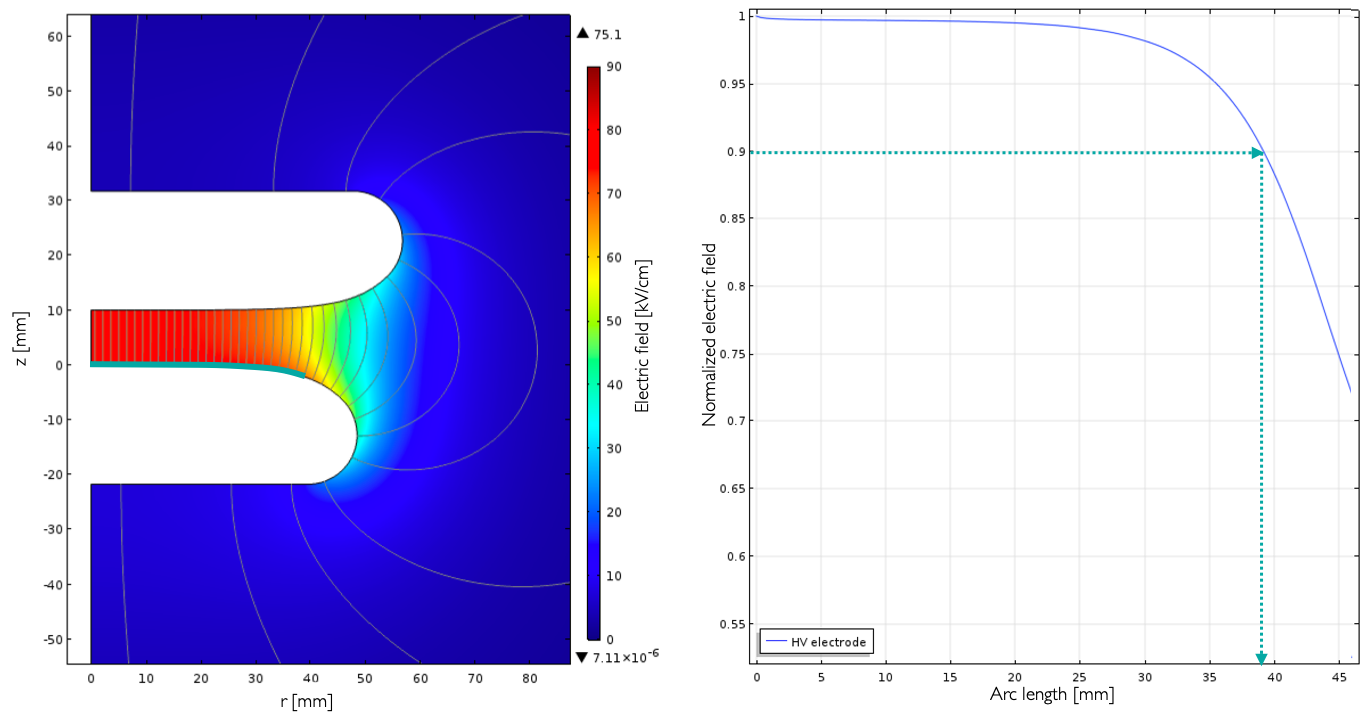}
\caption{Left:~the electric field between two Rogowski electrodes separated by 10~mm and enclosed in a distant grounded tube with -75~kV applied to the cathode. The teal line on the upper surface of the lower electrode illustrates the region of the cathode surface at this electrode spacing where the electric field magnitude is above 90\% of the maximum field. This line covers the stressed area of the electrode, and it shortens as the electrodes approach. Right:~plot of the electric field magnitude along the cathode surface for the geometry on the left normalized to the maximum electric field. The dotted teal arrows illustrate the determination of the stressed electrode radius, which contains electric fields above 90\% of the maximum field and bounds the stressed area.}
\label{fig:SEA}
\end{center} 
\end{figure}

A key quantity of interest is the stressed cathode (or electrode) area (SEA). Here, following other publications, it is defined as the area of the cathode surface with an electric field magnitude exceeding 90\% of the maximum electric field, as illustrated in figure~\ref{fig:SEA}. The radius corresponding to the boundary of the SEA was obtained from simulations in COMSOL Multiphysics v5.0$^{\circledR}$~\cite{comsolRef}. The area was then calculated from this radius by assuming that the Rogowski electrodes are perfectly flat within the region of interest. Note that as the electrode separation increases, the SEA increases as well, but the electric field becomes slightly less uniform.

To achieve a comparable state for measurements at each electrode separation, data at a given separation were usually collected during the day after the system was held at a stable pressure overnight. This resulted in more breakdown events measured at smaller separations since less time was needed to ramp to lower voltages. The location of the sparks between the electrodes varied between subsequent breakdowns and spanned the entire stressed surface of the electrodes. The data were acquired only when the fluid around the electrodes and the ceramic feedthrough was seen to be in a bubble-free state. The detector was not assembled in a clean room and specks of dust were sometimes visible on the electrode surfaces, usually near the beginning of each run. The dust moved away from the electrode surfaces due to liquid circulation, convective effects of breakdown, and bubbles. Note that the temperature inside the central chamber varied by 7~degrees in LAr and by 11~degrees in LXe throughout the measurements.

The purity monitor in LAr indicated an electron lifetime of 400~$\mu$s, which corresponds to 1~ppb oxygen equivalent impurities~\cite{buckley1989pm_grid_transparency,bakale1976impurity_attachment,Tvrznikova:2018pcz}. However, due to a purity monitor malfunction, this represents a lower limit on the electron lifetime. Purity measurements in LXe were not performed using the purity monitor due to technical difficulties. The composition of xenon used in Run~2 was instead measured with a residual gas analyzer and found to contain at most $\sim$200~ppm oxygen. In Run~3, the purity was measured using the scintillation properties of xenon~\cite{Chepel,ARNEODO2000147}, resulting in an electron lifetime of 2.2~$\mu$s, or 200~ppb oxygen equivalent impurities. This information is summarized in table~\ref{tab:Summary-of-data}.

\begin{table}
\begin{centering}
\begin{tabular}{llrrrrr}
\hline 
 & Liquid & Electrode & Area & Pressure & Electron & Impurity \tabularnewline
 &  & separation & range &  & lifetime & concentration \tabularnewline
 &  & {$[$mm$]$} & {$[$cm$^{2}]$} & {$[$bar$]$} & {$[\mu$s$]$} & {$[$ppb$]$} \tabularnewline
\hline 
Run 1 & Ar & 1-6 & 11.2-32.6 & 1.5, 2 & 400 & 1\tabularnewline
Run 2 & Xe & 1-5 & 11.2-28.7 & 2 & $\sim0.002$ & $\sim2\times10^5$\tabularnewline
Run 3 & Xe & 1-2 & 11.2-17.1 & 2 & 2.2 & 200\tabularnewline
\hline 
\end{tabular}
\par\end{centering}
\caption[Summary of data acquisitions used in analysis]{Summary of data acquisitions. The impurity concentration refers to the concentration of oxygen that would have an equivalent effect on electron lifetime or scintillation light. In Run~2 a residual gas analyzer was used to measure the concentration of oxygen in xenon directly and was converted to electron lifetime for convenience.
\label{tab:Summary-of-data}}
\end{table}

\subsection{Statistical and systematic errors}

Sufficient data were collected at each separation to suppress statistical errors, so systematic errors dominate the results presented in this work. Several sources of systematic error were considered. A 0.1~mm uncertainty was estimated for the electrode separation based on the mechanical error of zeroing the electrodes. Additionally, errors in the parallelism of the electrodes were considered. Our experiments with the visual perception of parallelism limit this uncertainty to 0.3$^{\circ}$ in each axis, giving a 0.4$^{\circ}$ systematic parallelism error. A 100~V uncertainty on the breakdown voltage was estimated based on the minimum power supply voltage increment. 

Errors from pressure fluctuations and conditioning effects were also taken into account. To estimate the systematic error caused by the variation of pressure, the data were divided into low- and high-pressure subsets for each electrode separation. The difference in the average breakdown value for the lower and higher pressure data sets was then taken as a systematic error on the breakdown voltage. Similarly, to account for conditioning effects, the data at each electrode separation were split into two groups based on the time of data collection. The difference in breakdown voltage for the earlier and later halves of data was treated as a systematic error. All the systematic errors were treated as uncorrected uncertainties in the analysis. The statistical and systematic errors of the LAr measurements are shown in table~\ref{tab:errors}.

\begin{table}
\begin{centering}
\begin{tabular}{r|>{\raggedleft}p{1.5cm}rr>{\raggedleft}p{1.3cm}>{\raggedleft}p{1.5cm}>{\raggedleft}p{1.7cm}>{\raggedleft}p{1.7cm}}
\hline 
$d$ & $\Delta$ separation & $\Delta$ tilt & $\Delta$ voltage & $\Delta$ pres-sure & $\Delta$ conditioning & Statistical error & Combined error\tabularnewline
{[}mm{]} & \% & \% & \% & \% & \% & \% & \%\tabularnewline
\hline 
1 & 10 & 5 & 0.9 & 3.8 & 2.1 & 1.2 & 12.1\tabularnewline
2 & 5 & 3 & 0.5 & 4.4 & 2.6 & 2.0 & 8.0\tabularnewline
3 & 3 & 2 & 0.3 & 4.5 & 3.1 & 1.5 & 6.9\tabularnewline
4 & 3 & 1 & 0.2 & 2.1 & 4.8 & 1.9 & 6.2\tabularnewline
5 & 2 & 1 & 0.2 & 2.1 & 4.3 & 2.9 & 6.0\tabularnewline
6 & 2 & 1 & 0.2 & 16.7 & 14.8 & 2.2 & 22.5\tabularnewline
\hline 
\end{tabular}
\par\end{centering}
\caption[Summary of systematic and statistical errors in LAr measurements]{Summary of systematic and statistical errors in LAr measurements. All errors are quoted as a percent error on the breakdown field. 
\label{tab:errors}}
\end{table}

\section{Results}

\subsection{Liquid argon}

\begin{figure}
\begin{center}
\includegraphics[width=100mm]{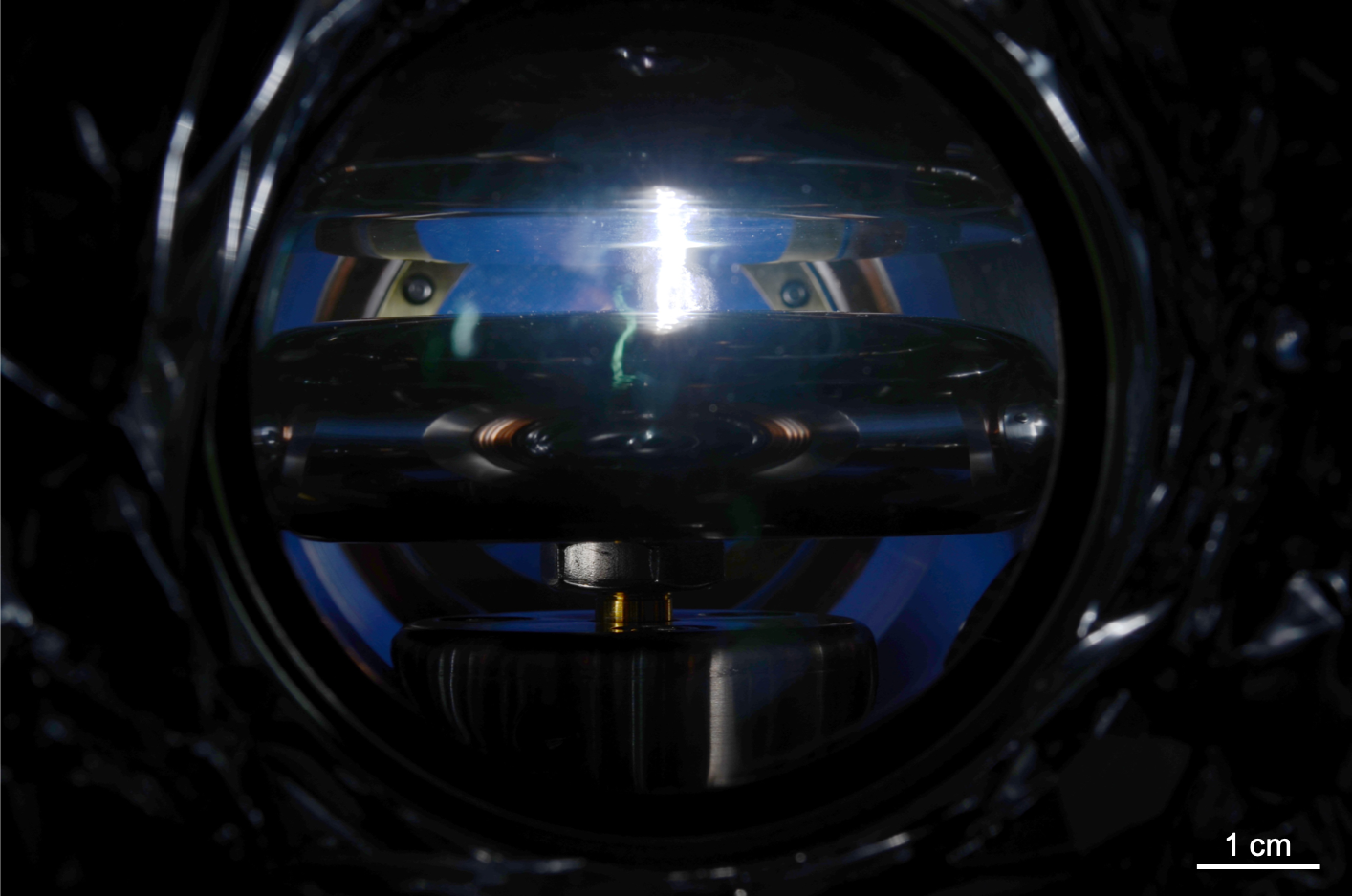}
\caption{Photograph of a spark in LAr for electrodes separated by 7~mm. All the observed flashes appeared to be initiated at the cathode. Breakdown data were only acquired in the absence of bubbles.}
\label{fig:LAr_spark}
\end{center} 
\end{figure}

\begin{figure}
\begin{center}
\includegraphics[width=140mm]{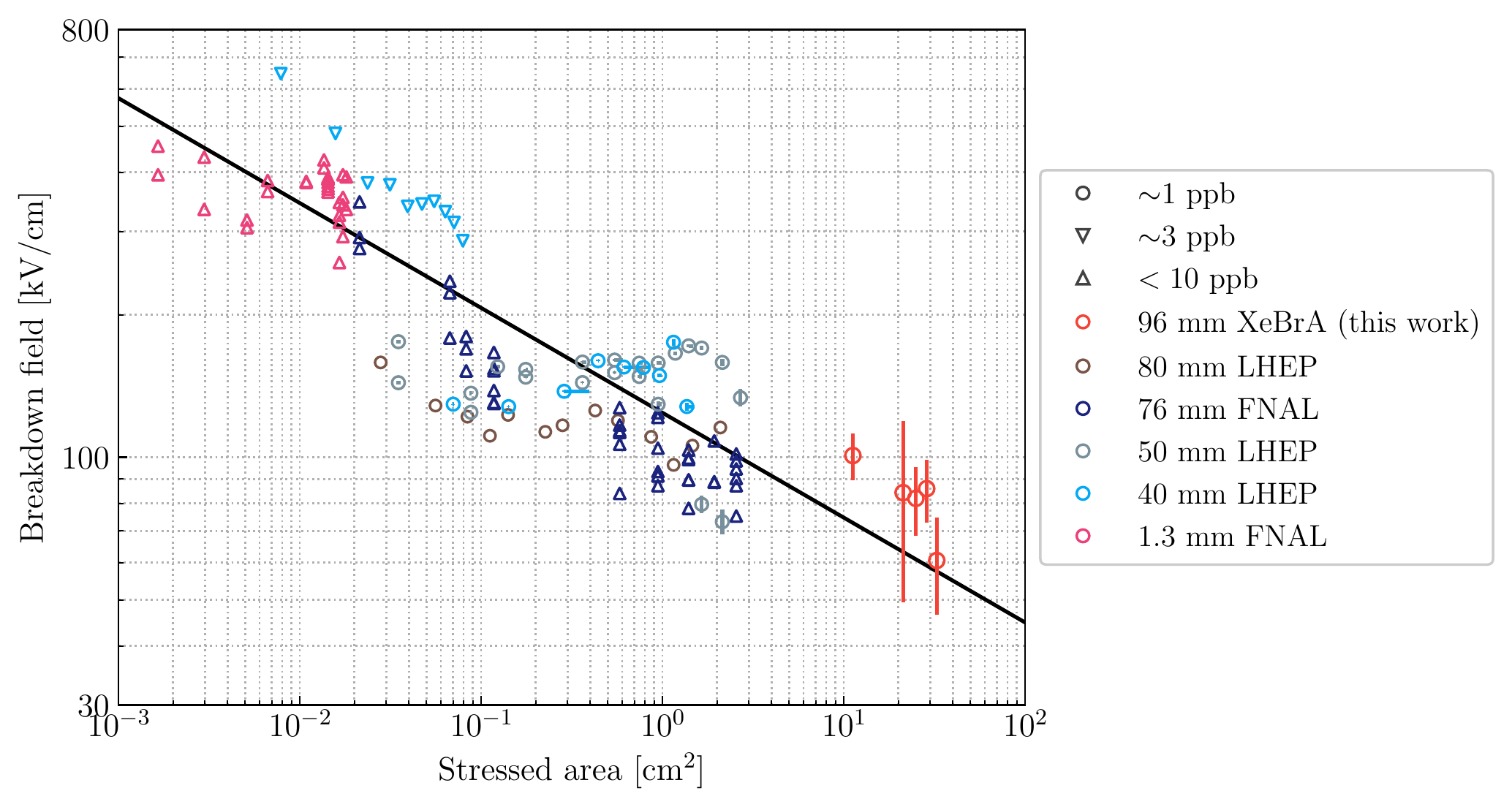}
\caption{A plot of breakdown field vs. 90\% stressed area of the cathode in LAr at $\sim1.5$~{bar}. The different colors indicate the cathode diameter used in each experiment, while the various shapes indicate LAr purity. The fit line corresponds to equation~(\ref{eq:AreaScalingGeneric}), $E_{M}=C\cdot (A/\mathrm{cm}^2)^{-b}$, where $C=124.26\pm0.09$~kV/cm and $b=0.2214\pm0.0002$ with $\chi^{2}=5\cdot10^{5}$ and 129 degrees of freedom (DOF). Data from XeBrA were collected at 1.5~bar. Data from the Fermi National Accelerator Laboratory (FNAL) were published in~\cite{Acciarri:2014ica} and were taken at 1.0-1.6~bar. Measurements from the Laboratory of High Energy Physics (LHEP) at the University of Bern were published in~\cite{Auger:2015xlo} and were taken at 1.2 bar. Note that electrodes with various geometries, dimensions, and separations (and hence maximal voltages) are included in this plot to study the area effect.}
\label{fig:LAr_area}
\end{center} 
\end{figure}

Figure~\ref{fig:LAr_spark} shows a picture of a spark in LAr. The characteristic behavior of the dielectric breakdown field as a function of stressed electrode area is shown in figure~\ref{fig:LAr_area}. This figure combines experimental data available in the literature with data obtained in XeBrA at 1.5~bar. All data in XeBrA were taken using the same electrodes by varying the separation from 1~to 6~mm, which changed the stressed area. Each data point from XeBrA is the mean value of all measurements from one run taken at the same electrode distance. Figure~\ref{fig:LAr_area} shows the breakdown field as a function of 90\% stressed cathode area, using data with impurity levels lower than 10~ppb available in the literature.

\begin{figure}
\begin{center}
\includegraphics[width=100mm]{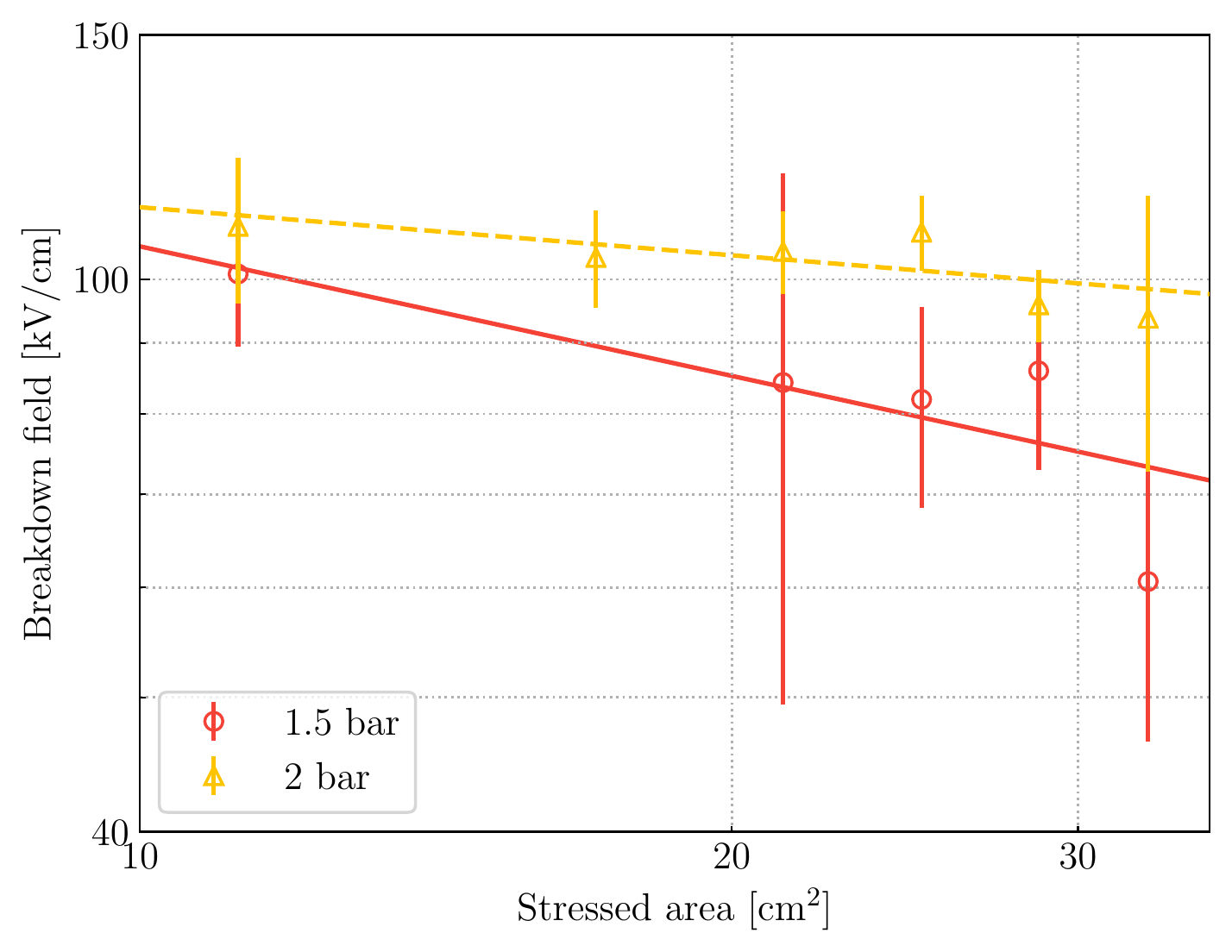}
\caption{Comparison of breakdown performance in LAr for data taken at 1.5~bar and 2~bar as a function of stressed electrode area. The fit at each pressure was performed to $E_{M}=C\cdot (A/\mathrm{cm}^2)^{-b}$. At 1.5~bar $C=216\pm103$~kV/cm and $b=0.31\pm0.16$ (solid red line) with $\chi^{2}=1.4$ and DOF = 2, while at 2~bar $C=147\pm54$~kV/cm and $b=0.11\pm0.12$ (dashed teal line) with $\chi^{2}=1.7$ and DOF~=~3.}
\label{fig:LAr_pressure}
\end{center} 
\end{figure}

Since data in LAr in XeBrA were collected at two different pressures, figure~\ref{fig:LAr_pressure} compares the breakdown performance at 1.5~bar and 2~bar. The data at each pressure were fit to equation~(\ref{eq:AreaScalingGeneric}). At 1.5~bar the slope was $b=0.31\pm0.16$ while at 2~bar $b=0.11\pm0.12$. Although not conclusive, the slope at 1.5~bar appears to be steeper than for 2~bar, consistent with the effects observed in LN and LHe~\cite{Gerhold1989,KAWASHIMA1974217,long2006high,Yoshino1982}. It is possible that this reflects a change in the propensity of bubble formation to act as a breakdown initiation mechanism. If this were the case, the relevant parameter is the difference between the absolute pressure and the vapor pressure of the fluid contacting the electrode surfaces, rather than simply the absolute pressure. An experiment designed to precisely measure and control the temperature between the electrodes could determine pressure effects definitively.

Additionally, a picoammeter measured the leakage current at the anode, but no apparent dependence on cathode voltage was observed. The leakage current was $<50$~fA in LAr.

\subsection{Liquid xenon}

\begin{figure}
\begin{center}
\includegraphics[width=100mm]{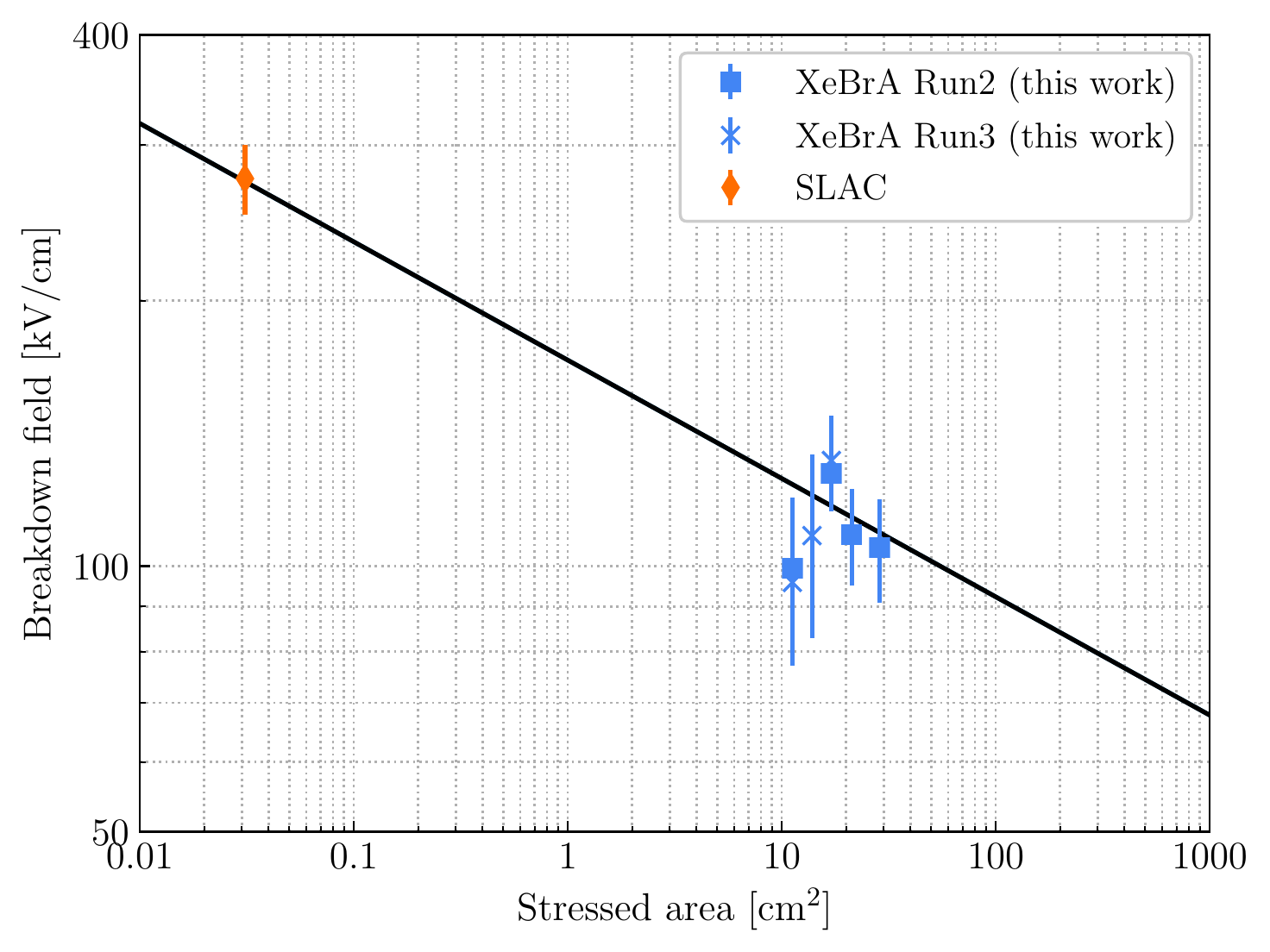}
\caption{A plot of breakdown field vs. 90\% stressed area of the cathode in LXe. The fit line corresponds to $E_{M}=C\cdot (A/\mathrm{cm}^2)^{-b}$ where $C=171\pm8$~kV/cm and $b=0.13\pm0.02$ with $\chi^{2}=6.5$ and DOF = 5. The SLAC result is taken from~\cite{Rebel:2014uia}. Data from XeBrA were taken at 2~bar. This presents a positive outlook for the LZ dark matter experiment~\cite{Mount:2017qzi}: The largest cathodic surface within the LZ detector is the cathode grid ring, with a simulated 90\% stressed area of 500~cm$^2$ and a maximum field of 26~kV/cm, while a maximum field of 29~kV/cm over an area of 150~cm$^2$ is simulated at the first reverse field ring. These values correspond to the LZ cathode design voltage of 100~kV and are $\sim$3x below the breakdown fields predicted by the above scaling.}
\label{fig:LXe_area}
\end{center} 
\end{figure}

Data collected in Runs~2 and 3 did not exhibit dependence on purity, so both acquisitions are included in the following analysis. The characteristic behavior of dielectric breakdown as a function of 90\% stressed electrode area is shown in figure~\ref{fig:LXe_area}. All data were taken using the same electrodes by varying the separation from 1~to 5~mm in Run~2 and from 1 to 2~mm in Run 3. A fit to equation~(\ref{eq:AreaScalingGeneric}), $E_{M}=C\cdot (A/\mathrm{cm}^2)^{-b}$, was performed where $C=171\pm8$~kV/cm and $b=0.13\pm0.02$. This fit includes a data point from SLAC National Accelerator Laboratory~\cite{Rebel:2014uia} as shown in the figure. The volume effect was also studied, see reference~\cite{Tvrznikova:2018pcz} for details.

A picoammeter measured the leakage current at the anode, but no apparent dependence on cathode voltage was observed. The leakage current was $<5$~fA in LXe.

\subsection{Comparison of liquid argon and liquid xenon data}

\begin{figure}
\begin{center}
\includegraphics[width=100mm]{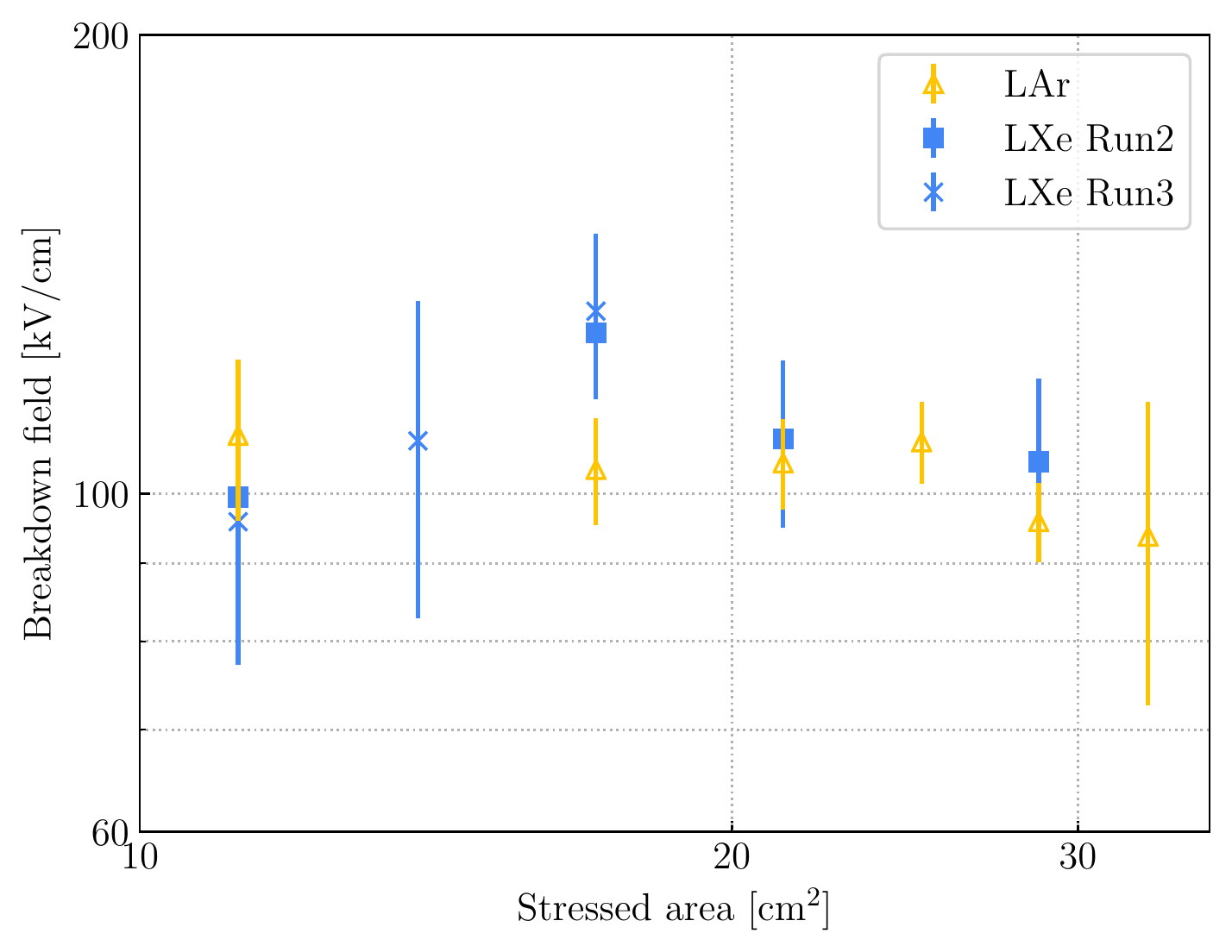}
\caption{Comparison of LAr and LXe data acquired in XeBrA at 2 bar.}
\label{fig:LAr_LXe_area}
\end{center} 
\end{figure}

The unique construction of XeBrA enables direct comparison of data acquired in LXe and LAr with the same geometry and electrodes. Figure~\ref{fig:LAr_LXe_area} shows all data obtained in XeBrA in LXe and LAr as a function of the 90\% stressed electrode area. There does not appear to be a significant difference between breakdown behavior in LXe and LAr.

\subsection{Modeling breakdown with a Weibull function}

\begin{figure}
\begin{center}
\includegraphics[width=100mm]{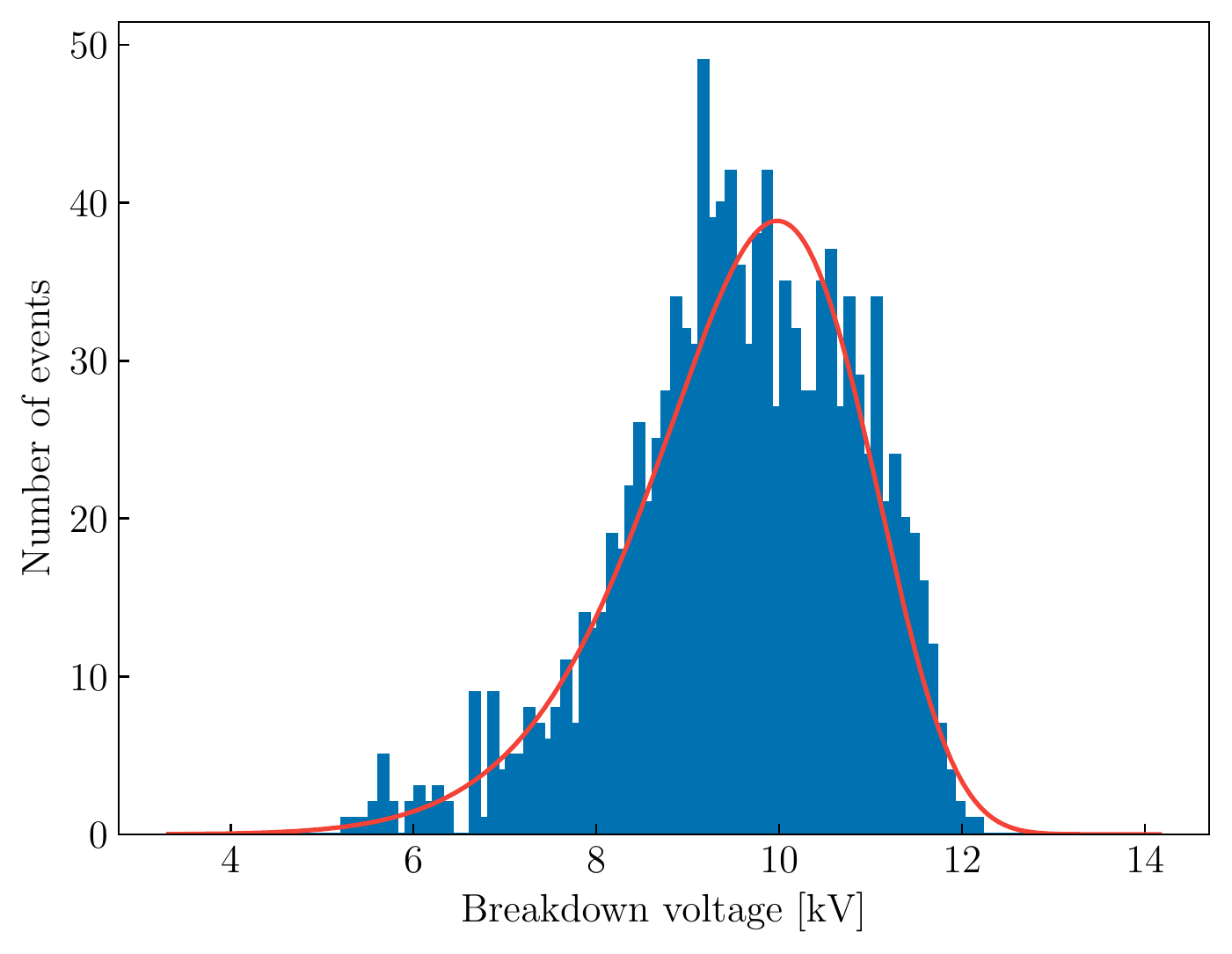}
\caption{Histogram of breakdown values for 1~mm electrode separation in LXe, corresponding to 11.2~cm$^2$ SEA. The solid red line shows a fit to the Weibull function with $k=9.2\pm0.2$ and $\lambda=10.10\pm0.03$ kV/cm. Error bars (not shown) are defined as $\sqrt N$, where $N$ is the number of events in each bin. The $\chi^{2}$ value of the fit is 99.8, with 63 degrees of freedom.}
\label{fig:Weibull_fit}
\end{center} 
\end{figure}

To test the weakest-link theory, the 2-parameter Weibull function of equation (\ref{eq:weibull_notime}) was fit to the data from LXe Run~3. The distributions for other data acquisitions appeared Weibull-like, but due to the small number of samples, the errors of the fit parameters were too large for a meaningful result. The resulting fit to data in LXe Run~3 from a 1~mm electrode separation is shown in figure~\ref{fig:Weibull_fit}. The values of the fitted parameters were $k=9.2\pm0.2$ and $\lambda=10.10\pm0.03$ kV/cm. A fit to a Gumbel distribution was also attempted but did not yield good agreement with data. 

Ideally, a fit to the breakdown distribution at any electrode separation from Run~3 in LXe would yield a similar value of the $k$ parameter. Therefore, data from other electrode separations in LXe Run~3 were also fit; for 1.4~mm separation the fit yielded $k=8.2\pm0.9$, and for 2~mm separation the fit yielded $k=12.8\pm1.8$. The value of $k$ can be compared to that obtained from the fit to the area scaling in figure~\ref{fig:LXe_area}, since $k=1/b$. The value of $k$ obtained from the area scaling is $k=7.7\pm1.2$. The $k$ values obtained from the Weibull fits are within 2~$\sigma$ of each other and of that obtained from the area scaling.

\section{Discussion}

The first data from XeBrA shed some light on the characteristics of HV breakdown in LAr and LXe. The dielectric strength of these two liquids has been measured with the same apparatus for the first time and shown to be very similar at the areas investigated. The measurements extend the area scaling power law previously observed in LAr to larger scales and suggest that a similar trend exists in LXe. The distribution of breakdown fields at a fixed gap in LXe is well described by a Weibull function and is consistent with the limited data available on area scaling in LXe. However, further studies are needed to confirm the validity of the Weibull hypothesis as an explanation for the stressed-area scaling effect. Since the current XeBrA configuration cannot accommodate larger electrodes, smaller electrodes can be installed to evaluate breakdown at a broader range of SEA and to explore breakdown statistics in LXe. A future study, with larger sample sizes and therefore more opportunity for statistical rigor, could confirm the applicability of the weakest-link theory. 

One explanation for the stressed area effect is that increasing the locally stored energy raises the probability that a weak initiating event grows into a breakdown event. This proposes that the area effect is simply a consequence of the correlation between the SEA and inter-electrode capacitance~\cite{cross1982,mazurek1987}. In XeBrA, the capacitive energy available to the electrodes is dominated by the capacitance of the HV supply cable. This provides an opportunity to test whether the available capacitive energy affects breakdown behavior by simply increasing the length of the cable. 

Many parameters that may affect dielectric breakdown in noble liquids can be explored with the XeBrA apparatus. Electrodes from various materials, with diverse coatings~\cite{1748-0221-9-07-P07023,SPSnEDM}, and with different surface finishes~\cite{highPressureWash,Bernard:1991du} can be studied. A promising direction to explore is whether the surface treatments that result in the dramatic reduction of electron emission from stainless steel wires also increase the dielectric strength of noble liquids between similarly-treated stainless steel electrodes~\cite{TOMAS201849}. Additionally, various impurities could be introduced to the system to investigate their effects on dielectric breakdown and electron lifetime. Further studies will also simultaneously include acquisition of signals from PMT, picoammeter, and charge-sensitive amplifier channels to study the onset of leakage currents and electroluminescence prior to a full breakdown in LAr and LXe. 

\section{Summary}
Dielectric breakdown behavior is a complex process influenced by many parameters that are often correlated in ways that make their effects difficult to distinguish. This difficulty is compounded when comparing results among studies because of the large number of uncontrolled and unreported parameters.

XeBrA is the first experiment to provide systematic studies of dielectric breakdown in LAr and LXe at stressed electrode areas greater than 3~cm$^2$. Results from XeBrA have further validated the existence of an area scaling effect for breakdown behavior in these liquids, but the HV breakdown in XeBrA in LAr occurred at slightly higher fields than what scaling predicts from previous LAr studies. The breakdown behavior of LAr and
LXe within the same experimental apparatus was comparable and no dependence of leakage current on the electric field was observed in either noble liquid. Preliminary data suggest that breakdown occurs at higher fields in LAr at higher pressures, but due to the substantial uncertainties in experimental conditions, no definitive conclusion regarding breakdown dependence on pressure is possible. 

This study also fit a breakdown distribution from LXe with a Weibull function to examine the applicability of the weakest-link theory. The parameter values obtained from the Weibull fits were roughly consistent with each other and with that obtained from the stressed area scaling but further data are needed to confirm this hypothesis.

\begin{acknowledgments}
The authors would like to thank Damian Goeldi, Igor Kreslo, Sarah Lockwitz, Wanchun Wei, Ian Smith, and Kareem Kazkaz for helpful correspondence. We gratefully acknowledge the technical and engineering staff at LBNL for their support. This work was supported by the Laboratory Directed Research and Development Program of Lawrence Berkeley National Laboratory under U.S. Department of Energy Contract No. DE-AC02-05CH11231. LLLN-JRNL-788151.

\end{acknowledgments}

\bibliographystyle{JHEP}
\bibliography{xebra}

\providecommand{\href}[2]{#2}\begingroup\raggedright\begin{thebibliography}{10}

\bibitem{Aalseth:2017fik}
{\scshape DarkSide} collaboration, C.~E. Aalseth et~al., \emph{{DarkSide-20k: A
  20 tonne two-phase LAr TPC for direct dark matter detection at LNGS}},
  \href{http://dx.doi.org/10.1140/epjp/i2018-11973-4}{\emph{Eur. Phys. J. Plus}
  {\bfseries 133} (2018) 131},
  [\href{https://arxiv.org/abs/1707.08145}{{\ttfamily 1707.08145}}].

\bibitem{Abi:2018dnh}
{\scshape DUNE} collaboration, B.~Abi et~al., \emph{{The DUNE Far Detector
  Interim Design Report Volume 1: Physics, Technology and Strategies}},
  \href{https://arxiv.org/abs/1807.10334}{{\ttfamily 1807.10334}}.

\bibitem{Aalbers:2016jon}
{\scshape DARWIN} collaboration, J.~Aalbers et~al., \emph{{DARWIN: towards the
  ultimate dark matter detector}},
  \href{http://dx.doi.org/10.1088/1475-7516/2016/11/017}{\emph{JCAP} {\bfseries
  1611} (2016) 017}, [\href{https://arxiv.org/abs/1606.07001}{{\ttfamily
  1606.07001}}].

\bibitem{Albert:2017hjq}
{\scshape nEXO} collaboration, J.~B. Albert et~al., \emph{{Sensitivity and
  Discovery Potential of nEXO to Neutrinoless Double Beta Decay}},
  \href{http://dx.doi.org/10.1103/PhysRevC.97.065503}{\emph{Phys. Rev.}
  {\bfseries C97} (2018) 065503},
  [\href{https://arxiv.org/abs/1710.05075}{{\ttfamily 1710.05075}}].

\bibitem{swan1960influence}
D.~Swan and T.~Lewis, \emph{Influence of electrode surface conditions on the
  electrical strength of liquefied gases},
  \href{http://dx.doi.org/10.1149/1.2427647}{\emph{Journal of the
  Electrochemical Society} {\bfseries 107} (1960) 180--185}.

\bibitem{swan_LAr}
D.~W. Swan and T.~J. Lewis, \emph{The influence of cathode and anode surfaces
  on the electric strength of liquid argon},
  \href{http://dx.doi.org/10.1088/0370-1328/78/3/311}{\emph{Proceedings of the
  Physical Society} {\bfseries 78} (1961) 448}.

\bibitem{Acciarri:2014ica}
{\scshape MicroBooNE} collaboration, R.~Acciarri et~al., \emph{{Liquid Argon
  Dielectric Breakdown Studies with the MicroBooNE Purification System}},
  \href{http://dx.doi.org/10.1088/1748-0221/9/11/P11001}{\emph{JINST}
  {\bfseries 9} (2014) P11001},
  [\href{https://arxiv.org/abs/1408.0264}{{\ttfamily 1408.0264}}].

\bibitem{Auger:2015xlo}
M.~Auger, A.~Blatter, A.~Ereditato, D.~Goeldi, S.~Janos, I.~Kreslo et~al.,
  \emph{{On the Electric Breakdown in Liquid Argon at Centimeter Scale}},
  \href{http://dx.doi.org/10.1088/1748-0221/11/03/P03017}{\emph{JINST}
  {\bfseries 11} (2016) P03017},
  [\href{https://arxiv.org/abs/1512.05968}{{\ttfamily 1512.05968}}].

\bibitem{weber1956nitrogen}
K.~H. Weber and H.~S. Endicott, \emph{Area effect and its extremal basis for
  the electric breakdown of transformer oil},
  \href{http://dx.doi.org/10.1109/AIEEPAS.1956.4499314}{\emph{Transactions of
  the American Institute of Electrical Engineers. Part III: Power Apparatus and
  Systems} {\bfseries 75} (Jan, 1956) 371--381}.

\bibitem{weber1955areaEffect}
K.~H. Weber and H.~S. Endicott, \emph{Area effect - and its extremal basis -
  for the electric breakdown of insulating oil},  in \emph{1955 Conference on
  electrical insulation}, pp.~13--15, Oct, 1955.
\newblock \href{http://dx.doi.org/10.1109/EIC.1955.7533319}{DOI}.

\bibitem{KAWASHIMA1974217}
A.~Kawashima, \emph{Electrode area effect on the electric breakdown of liquid
  nitrogen},
  \href{http://dx.doi.org/10.1016/0011-2275(74)90192-1}{\emph{Cryogenics}
  {\bfseries 14} (1974) 217 -- 219}.

\bibitem{goshima1995nitrogen}
H.~Goshima, N.~Hayakawa, M.~Hikita and K.~Uchida, \emph{Area and volume effects
  on breakdown strength in liquid nitrogen},
  \href{http://dx.doi.org/10.1109/94.395426}{\emph{IEEE Transactions on
  Dielectrics and Electrical Insulation} {\bfseries 2} (June, 1995) 376--384}.

\bibitem{GERHOLD1994579}
J.~Gerhold, M.~Hubmann and E.~Telser, \emph{Gap size effect on liquid helium
  breakdown},
  \href{http://dx.doi.org/10.1016/0011-2275(94)90183-X}{\emph{Cryogenics}
  {\bfseries 34} (1994) 579 -- 586}.

\bibitem{Gerhold1989}
J.~Gerhold, \emph{Breakdown phenomena in liquid helium},
  \href{http://dx.doi.org/10.1109/14.90264}{\emph{IEEE Transactions on
  Electrical Insulation} {\bfseries 24} (April, 1989) 155--166}.

\bibitem{goshime1995weibull}
H.~Goshima, N.~Hayakawa, M.~Hikita, H.~Okubo and K.~Uchida, \emph{Weibull
  statistical analysis of area and volume effects on the breakdown strength in
  liquid nitrogen}, \href{http://dx.doi.org/10.1109/94.395427}{\emph{IEEE
  Transactions on Dielectrics and Electrical Insulation} {\bfseries 2} (June,
  1995) 385--393}.

\bibitem{Gerhold_LiquidHelium}
J.~Gerhold, M.~Hubmann and E.~Telser, \emph{{About the size effect in LHe
  breakdown}},  in \emph{{ICDL'96. 12th International Conference on Conduction
  and Breakdown in Dielectric Liquids}}, pp.~324--328, July, 1996.
\newblock \href{http://dx.doi.org/10.1109/ICDL.1996.565499}{DOI}.

\bibitem{toya1981}
H.~Toya, N.~Ueno, T.~Okada and Y.~Murai, \emph{Statistical property of
  breakdown between metal electrodes in vacuum},
  \href{http://dx.doi.org/10.1109/TPAS.1981.316537}{\emph{IEEE Transactions on
  Power Apparatus and Systems} {\bfseries PAS-100} (April, 1981) 1932--1939}.

\bibitem{plasma_physics}
P.~Vanraes and A.~Bogaerts, \emph{Plasma physics of liquids-a focused review},
  \href{http://dx.doi.org/10.1063/1.5020511}{\emph{Applied Physics Reviews}
  {\bfseries 5} (2018) 031103}.

\bibitem{TOMAS201849}
A.~Tom{\'a}s, H.~M. Ara{\'u}jo, A.~J. Bailey, A.~Bayer, B.~Chen,
  B.~L{\'o}pez~Paredes et~al., \emph{Study and mitigation of spurious electron
  emission from cathodic wires in noble liquid time projection chambers},
  \href{http://dx.doi.org/10.1016/j.astropartphys.2018.07.001}{\emph{Astroparticle
  Physics} {\bfseries 103} (2018) 49 -- 61}.

\bibitem{Yoshino1982}
K.~Yoshino, K.~Ohseko, M.~Shiraishi, M.~Terauchi and Y.~Inuishi,
  \emph{Dielectric breakdown of cryogenic liquids in terms of pressure,
  polarity, pulse width and impurity}, {\emph{Journal of Electrostatics}
  {\bfseries 12} (1982) 305--314}.

\bibitem{long2006high}
J.~C. Long et~al., \emph{{High voltage test apparatus for a neutron EDM
  experiment and lower limit on the dielectric strength of liquid helium at
  large volumes}},  \href{https://arxiv.org/abs/physics/0603231}{{\ttfamily
  physics/0603231}}.

\bibitem{ikeda_movingOil}
M.~Ikeda, T.~Teranishi, M.~Honda and T.~Yanari, \emph{Breakdown characteristics
  of moving transformer oil},
  \href{http://dx.doi.org/10.1109/TPAS.1981.316952}{\emph{IEEE Transactions on
  Power Apparatus and Systems} {\bfseries PAS-100} (Feb, 1981) 921--928}.

\bibitem{blatter2014experimental}
A.~Blatter et~al., \emph{{Experimental study of electric breakdowns in liquid
  argon at centimeter scale}},
  \href{http://dx.doi.org/10.1088/1748-0221/9/04/P04006}{\emph{JINST}
  {\bfseries 9} (2014) P04006},
  [\href{https://arxiv.org/abs/1401.6693}{{\ttfamily 1401.6693}}].

\bibitem{schwenterly1974}
S.~W. Schwenterly et~al., \emph{Dielectric strength of liquid helium in
  millimeter gaps},  in \emph{Conference on Electrical Insulation Dielectric
  Phenomena - Annual Report 1974}, pp.~585--593, Oct, 1974.
\newblock \href{http://dx.doi.org/10.1109/CEIDP.1974.7735954}{DOI}.

\bibitem{streamer}
A.~{Sun}, C.~{Huo} and J.~{Zhuang}, \emph{Formation mechanism of streamer
  discharges in liquids: a review},
  \href{http://dx.doi.org/10.1049/hve.2016.0016}{\emph{High Voltage} {\bfseries
  1} (2016) 74--80}.

\bibitem{cross1982}
J.~D. Cross, \emph{A physical explanation of the effects of electrode area on
  the breakdown of liquid dielectrics},
  \href{http://dx.doi.org/10.1109/CEEJ.1982.6594620}{\emph{Canadian Electrical
  Engineering Journal} {\bfseries 7} (April, 1982) 28--30}.

\bibitem{mazurek1987}
B.~Mazurek and J.~D. Cross, \emph{An energy explanation of the area effect in
  electrical breakdown in vacuum},
  \href{http://dx.doi.org/10.1109/TEI.1987.299000}{\emph{IEEE Transactions on
  Electrical Insulation} {\bfseries EI-22} (June, 1987) 341--346}.

\bibitem{ballat}
J.~Ballat, D.~Konig and U.~Reininghaus, \emph{Spark conditioning procedures for
  vacuum interrupters in circuit breakers},
  \href{http://dx.doi.org/10.1109/14.231544}{\emph{IEEE Transactions on
  Electrical Insulation} {\bfseries 28} (Aug, 1993) 621--627}.

\bibitem{yasuokaelectrode}
T.~Yasuoka, K.~Kato and H.~Okubo, \emph{Electrode conditioning characteristics
  in vacuum based on discharge current}, {\emph{Proc. ISETS07} (2007) }.

\bibitem{Olivier_Helium}
C.~Olivier, \emph{Electrode conditioning effects in liquid helium},  in
  \emph{1982 IEEE International Conference on Electrical Insulation},
  pp.~268--271, June, 1982.
\newblock \href{http://dx.doi.org/10.1109/EIC.1982.7464486}{DOI}.

\bibitem{Coletti1982}
G.~L. G.~Coletti, P.~Girdinio and G.~Molinari, \emph{Some data on the ac
  breakdown of liquid helium in the millimetre gas range}, {\emph{Journal of
  Electrostatics} {\bfseries 12} (1982) 297--304}.

\bibitem{Weibull1939}
W.~Weibull, \emph{A Statistical Theory of the Strength of Materials}.
\newblock Generalstabens litografiska anstalts forlag, Stockholm, 1939.

\bibitem{Weibull1951}
W.~Weibull, \emph{A statistical distribution function of wide applicability},
  {\emph{J. Appl. Mech} {\bfseries 18} (1951) 293--297}.

\bibitem{goshima_LN}
H.~Goshima, N.~Hayakawa, M.~Hikita, K.~Uchida and H.~Okubo, \emph{{Statistical
  analysis of area and volume effects on breakdown voltage in liquid nitrogen
  using Weibull distribution}},  in \emph{Proceedings of 1994 IEEE
  International Symposium on Electrical Insulation}, pp.~430--433, June, 1994.
\newblock \href{http://dx.doi.org/10.1109/ELINSL.1994.401427}{DOI}.

\bibitem{haywaka_LN}
N.~Hayakawa, H.~Sakakibara, H.~Goshima, M.~Hikita and H.~Okubo, \emph{Breakdown
  mechanism of liquid nitrogen viewed from area and volume effects},
  \href{http://dx.doi.org/10.1109/94.590883}{\emph{IEEE Transactions on
  Dielectrics and Electrical Insulation} {\bfseries 4} (Feb, 1997) 127--134}.

\bibitem{gerhold_LN}
J.~Gerhold, M.~Hubmann and E.~Telser, \emph{{DC breakdown strength of liquid
  nitrogen under different voltage ramp conditions}},  in \emph{Proceedings of
  1999 IEEE 13th International Conference on Dielectric Liquids (ICDL'99) (Cat.
  No.99CH36213)}, pp.~445--448, July, 1999.
\newblock \href{http://dx.doi.org/10.1109/ICDL.1999.798968}{DOI}.

\bibitem{8124618}
W.~Yuan, T.~Wang, H.~Ni, M.~Gao, Y.~Ding, Y.~Li et~al., \emph{Weibull
  statistical analysis of size effects on the impulse breakdown strength in
  transformer oil},  in \emph{2017 IEEE 19th International Conference on
  Dielectric Liquids (ICDL)}, pp.~1--4, June, 2017.
\newblock \href{http://dx.doi.org/10.1109/ICDL.2017.8124618}{DOI}.

\bibitem{1996812}
H.~Goshima, H.~Sakakibara, N.~Hayakawa, M.~Hikita, K.~Uchida and H.~Okubo,
  \emph{High voltage insulation based on area and volume effects of cryogenic
  liquids for superconducting power apparatus},
  \href{http://dx.doi.org/10.1541/ieejpes1990.116.7_812}{\emph{IEEJ
  Transactions on Power and Energy} {\bfseries 116} (1996) 812--818}.

\bibitem{suehiro1996SuperfluidHe}
J.~Suehiro, K.~Ohno, T.~Takahashi, M.~Miyama and M.~Hara, \emph{Statistical
  characteristics of electrical breakdown in saturated superfluid helium},  in
  \emph{Conference Record of the ICDL '96. 12th International Conference on
  Conduction and Breakdown in Dielectric Liquids}, pp.~320--323, 1996.
\newblock \href{http://dx.doi.org/10.1109/ICDL.1996.565496}{DOI}.

\bibitem{alumina_breakdown}
B.~Block, Y.~Kim and D.~K. Shetty, \emph{Dielectric breakdown of
  polycrystalline alumina: A weakest-link failure analysis},
  \href{http://dx.doi.org/10.1111/jace.12492}{\emph{Journal of the American
  Ceramic Society} {\bfseries 96} (2013) 3430--3439}.

\bibitem{Ulhaq_polymer}
S.~Ul-Haq and G.~R.~G. Raju, \emph{Weibull statistical analysis of area effect
  on the breakdown strength in polymer films},  in \emph{Annual Report
  Conference on Electrical Insulation and Dielectric Phenomena}, pp.~518--521,
  Oct, 2002.
\newblock \href{http://dx.doi.org/10.1109/CEIDP.2002.1048847}{DOI}.

\bibitem{Hill_breakdown}
R.~M. Hill and L.~A. Dissado, \emph{Examination of the statistics of dielectric
  breakdown}, {\emph{Journal of Physics C: Solid State Physics} {\bfseries 16}
  (1983) 4447},
  [\href{https://arxiv.org/abs/http://stacks.iop.org/0022-3719/16/i=22/a=018}{{\ttfamily
  http://stacks.iop.org/0022-3719/16/i=22/a=018}}].

\bibitem{choulkov}
V.~V. Choulkov, \emph{Effect of electrode surface roughness on electrical
  breakdown in high voltage apparatus},
  \href{http://dx.doi.org/10.1109/TDEI.2005.1394020}{\emph{IEEE Transactions on
  Dielectrics and Electrical Insulation} {\bfseries 12} (Feb, 2005) 98--103}.

\bibitem{FisherTippett1928}
R.~A. Fisher and L.~H.~C. Tippett, \emph{Limiting forms of the frequency
  distribution of the largest or smallest member of a sample},
  \href{http://dx.doi.org/10.1017/S0305004100015681}{\emph{Proc. Cambridge
  Philos. Soc.} {\bfseries 24} (1928) 180}.

\bibitem{Gyorgyi2010}
G.~Gy{\"o}rgyi, N.~R. Moloney, K.~Ozog{\'a}ny, Z.~R{\'a}cz and M.~Droz,
  \emph{Renormalization-group theory for finite-size scaling in extreme
  statistics}, \href{http://dx.doi.org/10.1103/PhysRevE.81.041135}{\emph{Phys.
  Rev. E.} {\bfseries 81} (2010) 041135}.

\bibitem{Coles2001}
S.~Coles, \emph{An introduction to modeling of extreme values}.
\newblock Springer, London, 1~ed., 2001,
  \href{http://dx.doi.org/10.1007/978-1-4471-3675-0}{10.1007/978-1-4471-3675-0}.

\bibitem{Kotz2000}
S.~Kotz and S.~Nadarajah, \emph{Extreme value distributions}.
\newblock World Scientific, 2000,
  \href{http://dx.doi.org/10.1142/p191}{10.1142/p191}.

\bibitem{NevesAlves2008}
C.~Neves and M.~I. Fraga~Alves, \emph{Testing extreme value conditions- an
  overview and recent approaches}, {\emph{Rev. Stat.} {\bfseries 6} (2008)
  83--100}.

\bibitem{Gerhold1998Properties}
J.~Gerhold, \emph{Properties of cryogenic insulants},
  \href{http://dx.doi.org/10.1016/S0011-2275(98)00094-0}{\emph{Cryogenics}
  {\bfseries 38} (1998) 1063 -- 1081}.

\bibitem{rogowski}
N.~G. Trinh, \emph{Electrode design for testing in uniform field gaps},
  \href{http://dx.doi.org/10.1109/TPAS.1980.319754}{\emph{IEEE Transactions on
  Power Apparatus and Systems} {\bfseries PAS-99} (May, 1980) 1235--1242}.

\bibitem{Rogowski1923}
W.~Rogowski, \emph{Die elektrische festigkeit am rande des
  plattenkondensators},
  \href{http://dx.doi.org/10.1007/BF01656573}{\emph{Archiv f{\"u}r
  Elektrotechnik} {\bfseries 12} (Jan, 1923) 1--15}.

\bibitem{doi:10.1063/1.1724850}
R.~L. Amey and R.~H. Cole, \emph{Dielectric constants of liquefied noble gases
  and methane}, \href{http://dx.doi.org/10.1063/1.1724850}{\emph{The Journal of
  Chemical Physics} {\bfseries 40} (1964) 146--148}.

\bibitem{doi:10.1139/p70-033}
J.~Marcoux, \emph{{Dielectric constants and indices of refraction of Xe, Kr,
  and Ar}}, \href{http://dx.doi.org/10.1139/p70-033}{\emph{Canadian Journal of
  Physics} {\bfseries 48} (1970) 244--245}.

\bibitem{doi:10.1142/5113}
Y.~Suzuki, M.~Nakahata, Y.~Koshio and S.~Moriyama, \emph{Technique and
  Application of Xenon Detectors}.
\newblock World Scientific, 2003,
  \href{http://dx.doi.org/10.1142/5113}{10.1142/5113}.

\bibitem{SAWADA2003449}
R.~Sawada, J.~Kikuchi, E.~Shibamura, M.~Yamashita and T.~Yoshimura,
  \emph{Capacitive level meter for liquid rare gases},
  \href{http://dx.doi.org/10.1016/S0011-2275(03)00100-0}{\emph{Cryogenics}
  {\bfseries 43} (2003) 449 -- 450}.

\bibitem{bettini1991pm_initial}
A.~Bettini et~al., \emph{A study of the factors affecting the electron lifetime
  in ultra-pure liquid argon},
  \href{http://dx.doi.org/10.1016/0168-9002(91)90532-U}{\emph{Nuclear
  Instruments and Methods in Physics Research Section A} {\bfseries 305} (1991)
  177 -- 186}.

\bibitem{carugno1990pm_formula}
G.~Carugno, B.~Dainese, F.~Pietropaolo and F.~Ptohos, \emph{Electron lifetime
  detector for liquid argon},
  \href{http://dx.doi.org/10.1016/0168-9002(90)90176-7}{\emph{Nuclear
  Instruments and Methods in Physics Research Section A} {\bfseries 292} (1990)
  580 -- 584}.

\bibitem{Tvrznikova:2018pcz}
L.~Tvrznikova, \emph{{Sub-GeV Dark Matter Searches and Electric Field Studies
  for the LUX and LZ Experiments}}.
\newblock PhD thesis, Yale University, 2019.
\newblock \href{https://arxiv.org/abs/1904.08979}{{\ttfamily 1904.08979}}.

\bibitem{comsolRef}
``{\textsc{comsol} Multiphysics\textsuperscript{\textregistered}}.''
  {\url{http://www.comsol.com}}.

\bibitem{buckley1989pm_grid_transparency}
E.~Buckley et~al., \emph{A study of ionization electrons drifting over large
  distances in liquid argon},
  \href{http://dx.doi.org/10.1016/0168-9002(89)90710-9}{\emph{Nuclear
  Instruments and Methods in Physics Research Section A} {\bfseries 275} (1989)
  364 -- 372}.

\bibitem{bakale1976impurity_attachment}
G.~Bakale, U.~Sowada and W.~F. Schmidt, \emph{Effect of an electric field on
  electron attachment to sulfur hexafluoride, nitrous oxide, and molecular
  oxygen in liquid argon and xenon},
  \href{http://dx.doi.org/10.1021/j100564a006}{\emph{The Journal of Physical
  Chemistry} {\bfseries 80} (1976) 2556--2559}.

\bibitem{Chepel}
V.~Chepel and H.~Araujo, \emph{{Liquid noble gas detectors for low energy
  particle physics}},
  \href{http://dx.doi.org/10.1088/1748-0221/8/04/R04001}{\emph{JINST}
  {\bfseries 8} (2013) R04001},
  [\href{https://arxiv.org/abs/1207.2292}{{\ttfamily 1207.2292}}].

\bibitem{ARNEODO2000147}
F.~Arneodo et~al., \emph{Scintillation efficiency of nuclear recoil in liquid
  xenon},
  \href{http://dx.doi.org/https://doi.org/10.1016/S0168-9002(99)01300-5}{\emph{Nuclear
  Instruments and Methods in Physics Research Section A} {\bfseries 449} (2000)
  147 -- 157}.

\bibitem{Rebel:2014uia}
B.~Rebel et~al., \emph{{High Voltage in Noble Liquids for High Energy
  Physics}},
  \href{http://dx.doi.org/10.1088/1748-0221/9/08/T08004}{\emph{JINST}
  {\bfseries 9} (2014) T08004},
  [\href{https://arxiv.org/abs/1403.3613}{{\ttfamily 1403.3613}}].

\bibitem{Mount:2017qzi}
{\scshape LZ} collaboration, B.~J. Mount et~al., \emph{{LUX-ZEPLIN (LZ)
  Technical Design Report}},
  \href{https://arxiv.org/abs/1703.09144}{{\ttfamily 1703.09144}}.

\bibitem{1748-0221-9-07-P07023}
M.~Auger et~al., \emph{A method to suppress dielectric breakdowns in liquid
  argon ionization detectors for cathode to ground distances of several
  millimeters}, {\emph{Journal of Instrumentation} {\bfseries 9} (2014)
  P07023},
  [\href{https://arxiv.org/abs/http://stacks.iop.org/1748-0221/9/i=07/a=P07023}{{\ttfamily
  http://stacks.iop.org/1748-0221/9/i=07/a=P07023}}].

\bibitem{SPSnEDM}
T.~M. Ito et~al., \emph{An apparatus for studying electrical breakdown in
  liquid helium at 0.4 k and testing electrode materials for the neutron
  electric dipole moment experiment at the spallation neutron source},
  \href{http://dx.doi.org/10.1063/1.4946896}{\emph{Review of Scientific
  Instruments} {\bfseries 87} (2016) 045113}.

\bibitem{highPressureWash}
R.~Ready et~al., \emph{{Surface preparation of high voltage electrodes for the
  Ra EDM experiment}}, {\emph{In preparation} (2019) }.

\bibitem{Bernard:1991du}
P.~Bernard, D.~Bloess, W.~Hartung, C.~Hauviller, W.~Weingarten, P.~Bosland
  et~al., \emph{{Superconducting niobium sputter coated copper cavities at
  1500-MHz}}, {\emph{Part. Accel.} {\bfseries 40} (1992) 487--496}.

\end{thebibliography}\endgroup

\end{document}